\documentclass{aa}
\usepackage{graphics,epsfig,times}

\usepackage{natbib}
\bibpunct{(}{)}{;}{a}{}{,}

\newcommand{\ergcm}[1]{$\times 10^{#1}$ erg cm$^{-2}$ s$^{-1}$}

\newcommand{\ergs}[1]{$\times 10^{#1}$ erg s$^{-1}$}
\newcommand{\oergs}[1]{$10^{#1}$ erg s$^{-1}$}
\newcommand{\hcm}[1]{$\times 10^{#1}$ cm$^{-2}$}
\newcommand{\ohcm}[1]{$10^{#1}$ cm$^{-2}$}
\newcommand{\expo}[1]{$\times 10^{#1}$}
\newcommand{\oexpo}[1]{$10^{#1}$}
\newcommand{\nh}{N$_{\rm H}$}

\newcommand{\Halp}{H${\alpha}$}
\newcommand{\ltsima}{$\buildrel < \over \sim$}
\newcommand{\lsim}{\lower.5ex\hbox{\ltsima}}
\newcommand{\gtsima}{$\buildrel > \over \sim$}
\newcommand{\gsim}{\lower.5ex\hbox{\gtsima}}

\newcommand{\ma}{\hbox{\object{$[$MA93$]$\,1393}}}
\newcommand{\rxs}{\hbox{\object{RX\,J0103.6$-$7201}}}
\newcommand{\rxl}{\hbox{\object{RX\,J0146.9$+$6121}}}

\newcommand{\exo}{\hbox{\object{EXO\,053109$-$6609.2}}}

\begin{document}
 
\title{Discovery of 1323 s pulsations from \rxs: the longest period X-ray pulsar in the SMC
       \thanks{Based on observations with 
               XMM-Newton, an ESA Science Mission with instruments and contributions 
               directly funded by ESA Member states and the USA (NASA)}}
 
\author{F.~Haberl \and W.~Pietsch}

\titlerunning{Discovery of 1323 s pulsations from \rxs}
\authorrunning{Haberl et al.}
 
\offprints{F. Haberl, \email{fwh@mpe.mpg.de}}
 
\institute{Max-Planck-Institut f\"ur extraterrestrische Physik,
           Giessenbachstra{\ss}e, 85748 Garching, Germany
	   }
 
\date{Received 2 December 2004 / Accepted 16 March 2005}
 
\abstract{
XMM-Newton archival observations of the Be/X-ray binary candidate \rxs\ revealed pulsations
with a period of $\sim$1323 s. This makes \rxs\ the X-ray pulsar with the longest period
known in the Small Magellanic Cloud (SMC).
More than 150 X-ray observations of \rxs\ by ROSAT, Chandra and XMM-Newton show  
flux variations by a factor of 50 on time scales of days to years. Using the accurate positions 
obtained from ACIS-I images the optical counterpart is identified with a V = 14.6 mag emission line 
star.
EPIC spectra of \rxs\ above 1 keV are consistent with 
an absorbed power-law with column density between (6$-$9)\hcm{21}, except during one observation 
when an extraordinary high value of 1.1\hcm{23} was measured which strongly attenuated the power-law 
emission below 3 keV. A soft excess between 0.5 and 1.0 keV is evident in the spectra which becomes 
best visible in the highly absorbed spectrum. The soft component can be reproduced by a thermal 
plasma emission model with its luminosity strongly correlated with
the total intrinsic source luminosity. Including results from three other SMC Be/X-ray binaries extends the 
linear correlation over three orders of magnitude in source intensity, strongly suggesting that
the same mechanism is responsible for the generation of the soft emission.

}
 
\maketitle
 
\section{Introduction}

High mass X-ray binaries (HMXBs) are binary systems consisting of a compact stellar object 
accreting matter from a high mass early type star. In most cases the detection of pulsations 
in the X-ray flux with periods in the range of 0.07~s to 1500~s reveals a spinning neutron 
star orbiting an OB supergiant or a Be star. The large population of HMXBs makes the SMC 
unique among the galaxies of the Local Group. First, the absolute number of known HMXBs
is similar to that in the Milky Way, despite the comparatively small mass of the SMC and 
second, while the ratio of
supergiant to Be binaries in the Milky Way is 1/3.5 as found from a detailed literature search, 
the vast majority of HMXBs in the SMC has a Be star counterpart.

Currently 48 X-ray pulsars are known in the SMC \citep{2005MNRAS.356..502C}, from which
45 are most likely in HMXB systems as suggested by their X-ray properties \citep{2004A&A...414..667H}. 
Optical identifications confirmed 33 Be counterparts while only one supergiant system (SMC\,X-1) is known.
Additional candidates for Be/X-ray binaries were found by \citet{2000A&A...359..573H} from correlations 
of ROSAT catalogues of SMC sources with lists of \Halp\ emission line objects 
\citep[][hereafter MA93]{1993A&AS..102..451M} and by \citet{2003A&A...403..901S}.

One of the Be/X-ray binary candidates is \rxs, its ROSAT position (with an uncertainty of 4\arcsec)
is consistent with the emission line object \ma\ \citep{2000A&A...359..573H}. This hard ROSAT source 
was detected in PSPC \citep{2000A&AS..142...41H} and HRI \citep{2000A&AS..147...75S} pointed 
observations of the SMC. From three ROSAT observations a factor of three flux variability was seen
on time scales of years. A first analysis of early XMM-Newton data was performed by 
\citet{2003A&A...403..901S}. \rxs\ is located close to the supernova remnant 1E\,0102$-$72.3 in 
the North-East of the SMC which is frequently observed as calibration source for the 
instruments on-board XMM-Newton and Chandra. About 150 Chandra observations were performed 
with \rxs\ located on various ACIS CCD chips which are readily available from the Chandra archive. 
In addition twelve XMM-Newton observations with the EPIC instruments 
in various readout modes makes \rxs\ the most frequently observed extragalactic X-ray binary.
We analyzed the temporal and spectral properties of \rxs\ and present the results in this paper.

\section{Data analysis}

\begin{figure*}
\hbox{\resizebox{8.2cm}{!}{\includegraphics[angle=-90,clip=]{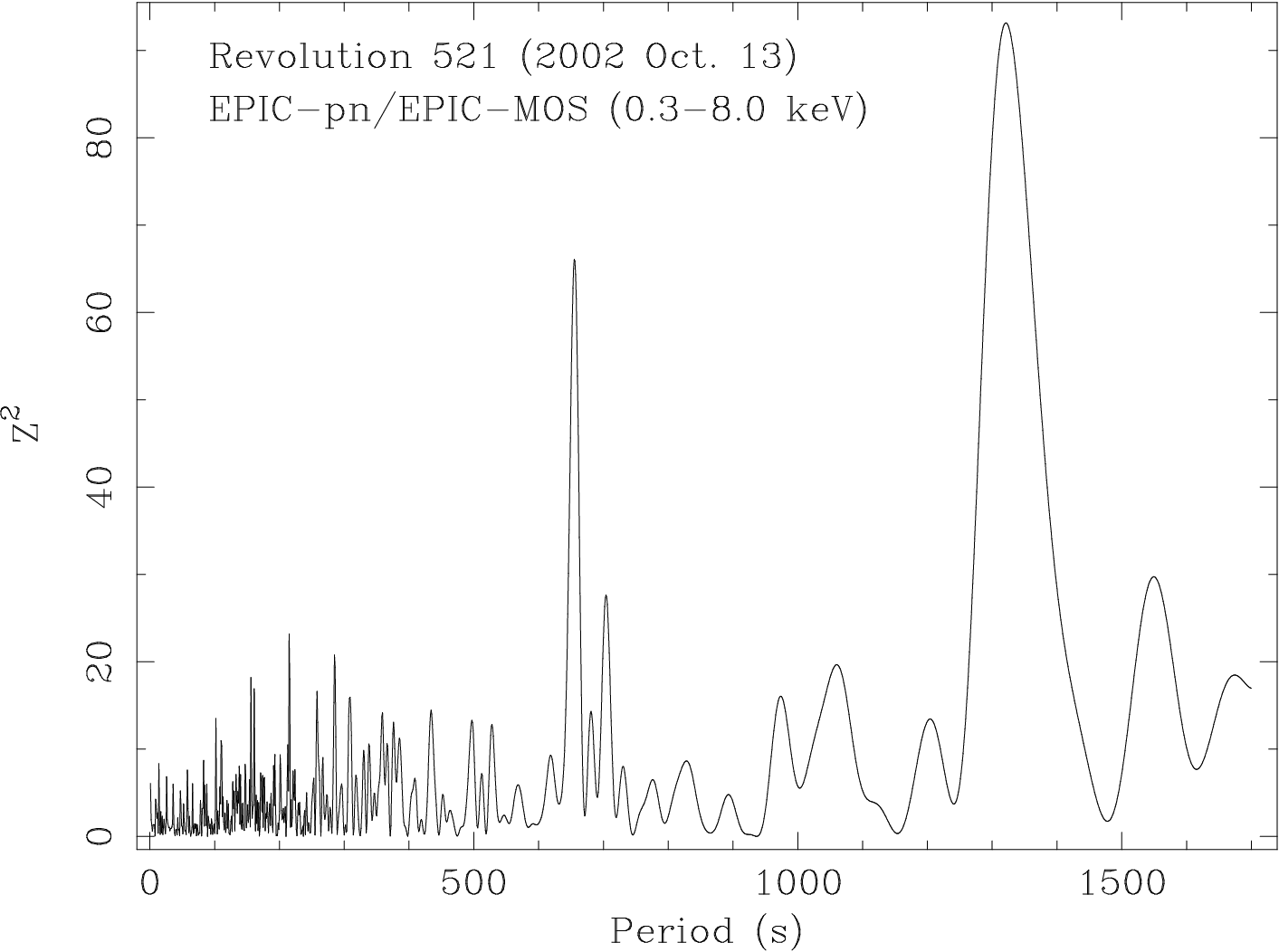}}
\hspace{4mm}
\resizebox{8.2cm}{!}{\includegraphics[angle=-90,clip=]{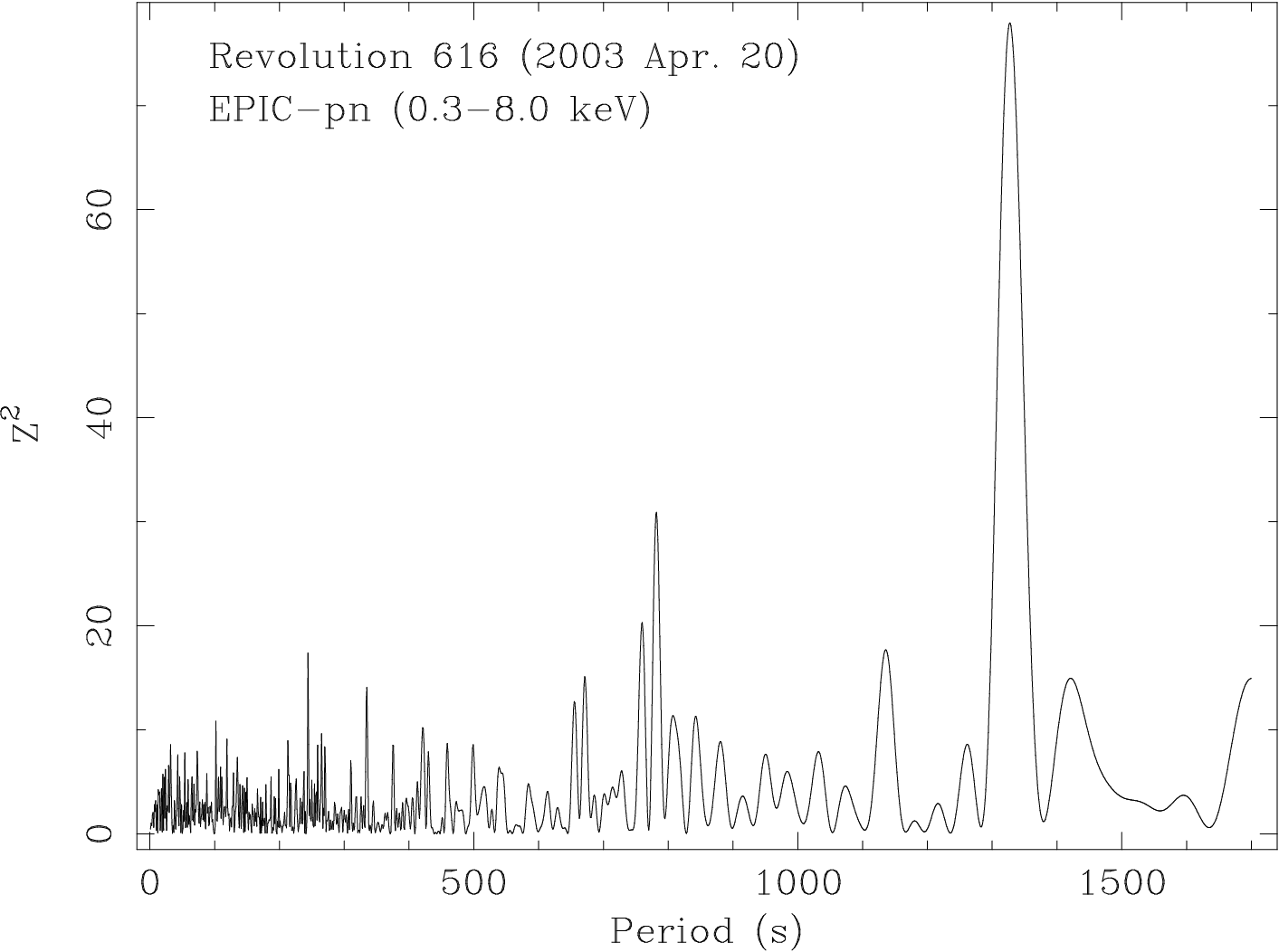}}}
\vspace{3mm}
\hbox{\resizebox{8.2cm}{!}{\includegraphics[angle=-90,clip=]{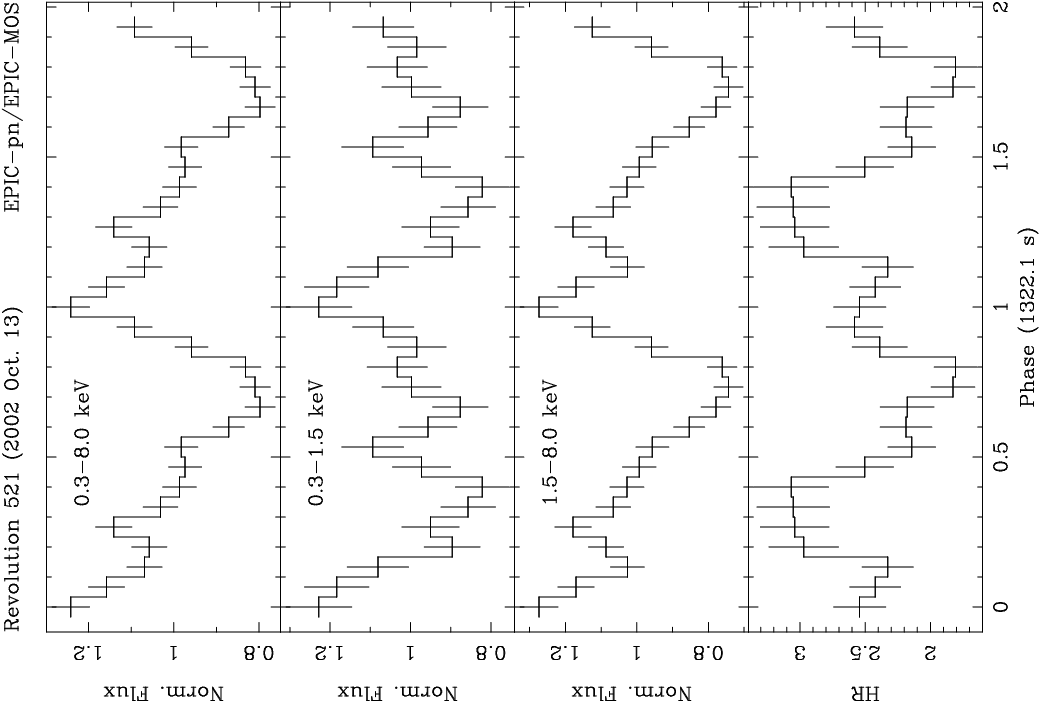}}
\hspace{5mm}
\resizebox{8.2cm}{!}{\includegraphics[angle=-90,clip=]{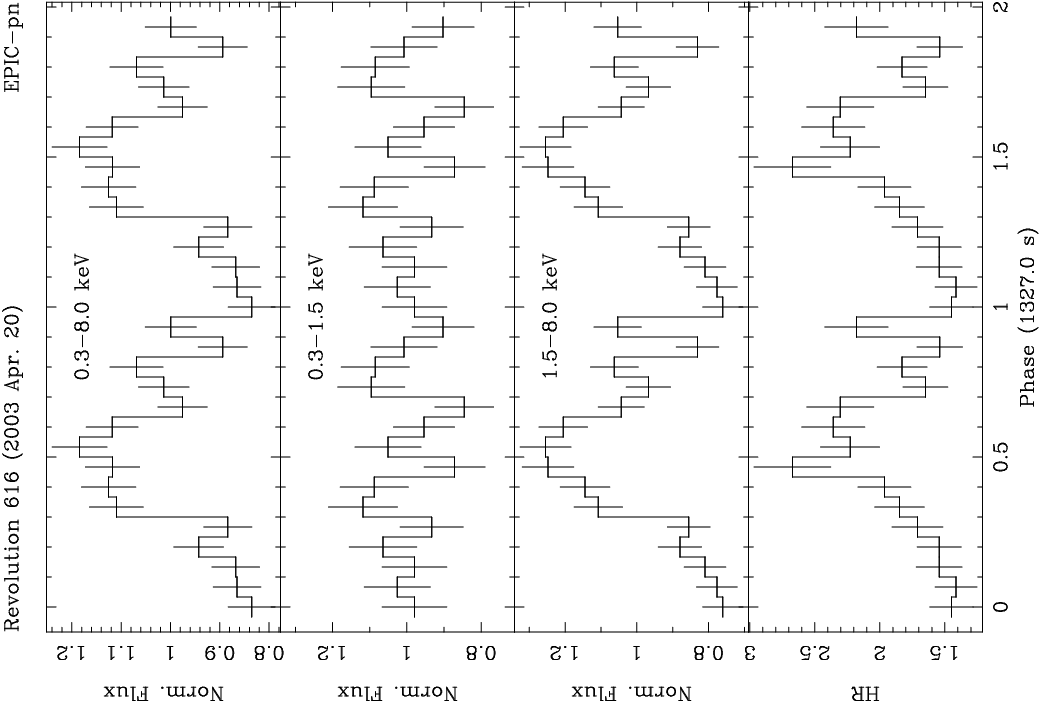}}}
\caption{Periodograms of the Z$^2_1$ test from two XMM-Newton observations of \rxs\ (top).
Corresponding pulse profiles folded with the most likely period for different energy bands, together with hardness
ratios (HR) are shown in the bottom panels.
For sinusoidal variations pulsed fractions in the 0.3$-$8.0 keV band 
are 15.7$\pm$1.7\% (rev. 521) and 13.2$\pm$2.2\% (rev. 616).}
\label{epic-z2}
\end{figure*}

The Chandra ACIS calibration observations of 1E\,0102$-$72.3 are typically of 8 ks exposure
(five observations have 20 ks) and were used to determine an accurate position (Sect. 2.4)
and to investigate the flux history of \rxs\ (Sect. 2.3). The twelve XMM-Newton EPIC observations 
performed until 2004, Oct. 15 have longer exposures of 20$-$30 ks (only one exception with 13 ks). 
The EPIC-pn \citep{2001A&A...365L..18S} and EPIC-MOS \citep{2001A&A...365L..27T} data were,
therefore, used for a spectral and temporal analysis, in particular to search for X-ray 
pulsations. During five of the EPIC observations (here as observation 
a single pointing within a satellite revolution is defined) two different CCD 
readout modes were used for the pn camera (switching between FF: full frame, LW: large window 
and SW: small window mode) while the MOS cameras were operated in LW mode during the whole
observation (only during revolution 433 and 447 part of the time was spent in timing mode).
Different optical light blocking filters (thin, medium and thick)
were chosen in different revolutions (see also Table~\ref{fit-spectra}). 
During revolutions 616 and 803 the MOS LW mode field of view did not cover \rxs.
Also the MOS timing mode observations cannot be used.
The data were processed using the XMM-Newton analysis package SAS 6.0.0 to 
produce the photon event files and binned data products such as images and spectra.
Events for spectral and temporal analysis of \rxs\ were 
extracted from circular regions (radius 20\arcsec) around the source position
and from nearby source-free regions for background spectra. Photon arrival times were corrected 
to the solar system barycentre. For the purpose of searching for relatively long
periods of seconds to tens of minutes as observed from Be/X-ray binaries the EPIC-pn data from one
satellite revolution with different readout modes were merged. MOS and pn data were combined
for the common time intervals, i.e. MOS data at the beginning of an observation when the pn camera
was still calculating the offset map (and also when the readout mode was changed) were disregarded.

\subsection{X-ray pulsations}

\begin{figure}
\resizebox{8.0cm}{!}{\includegraphics[angle=-90,clip=]{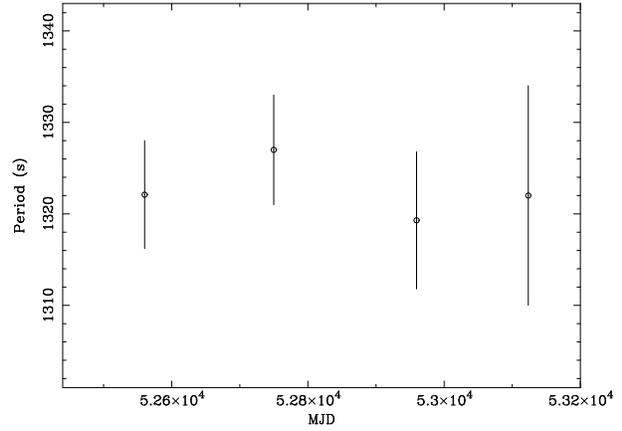}}
\caption{Pulse period history of \rxs\ over 1.5 years. The pulse periods of 
1322.1$\pm$5.9, 1327.0$\pm$6.0, 1319.3$\pm$7.5 and 1322$\pm$12 were derived from the EPIC data from revolutions
521, 616, 721 and 803, respectively.}
\label{period-hist}
\end{figure}

We performed a timing analysis of the EPIC data of \rxs, searching for pulsations in the broad 
energy band (0.3$-$8.0 keV) using the Rayleigh Z$^2_1$ technique in the range 1$-$3000 s. 
Strong peaks in the probability density function were found from the observations in revolutions
521 (pn SW+LW and MOS1/2 LW), 616 (pn SW+LW), 721 (pn FF and MOS1/2 LW) and 803 (pn LW) 
with Z$^2_1$ of 93.2, 77.9, 68.4 and 29.9, respectively. 
The maximum Z$^2_1$ of 93.2 derived for revolution 521 corresponds 
to a period detection confidence of 1.0$-$1.3\expo{-16} (disregarding the detection in other observations).
Fig.~\ref{epic-z2} shows two example periodograms (top) together with the 
EPIC X-ray light curves folded with the most likely period value derived from the 
Z$^2$ test. To investigate energy dependencies of the pulse profile, light curves in the 
0.3$-$1.5 keV (soft), 1.5$-$8.0 keV (hard) and the total 0.3$-$8.0 keV energy bands were created.
Hardness ratios derived by dividing the pulse profiles in the hard and soft bands show
significant variations caused by weaker pulse modulations in the soft band compared to the hard band.
This is particularly evident during the revolution 616 observation where no significant modulation
was seen in the soft band.
Period errors (1$\sigma$) were determined by $\chi^2$ fitting of a sine wave to the total 
light curves with 80 s binning, taking into account long term trends visible in the light curves
by adding polynomial components to the sine model light curve. We also tested the inclusion of 
a second harmonic component as might be suggested by the analysis shown in Fig.~\ref{epic-z2}. This
did not improve the results because the major uncertainties in the determination of the long pulse period
arise from the superposed long term trends.
The pulse period history for the four observations is shown in Fig.~\ref{period-hist}.
No significant change in the pulse period is visible consistent with a constant period of 
(1323.2 $\pm$ 5.8) s. However, the large errors are still compatible with large spin-up or 
spin-down with up to $\pm$13 s year$^{-1}$.

\begin{figure*}[ht]
\hbox{\resizebox{7.8cm}{!}{\includegraphics[angle=-90,clip=]{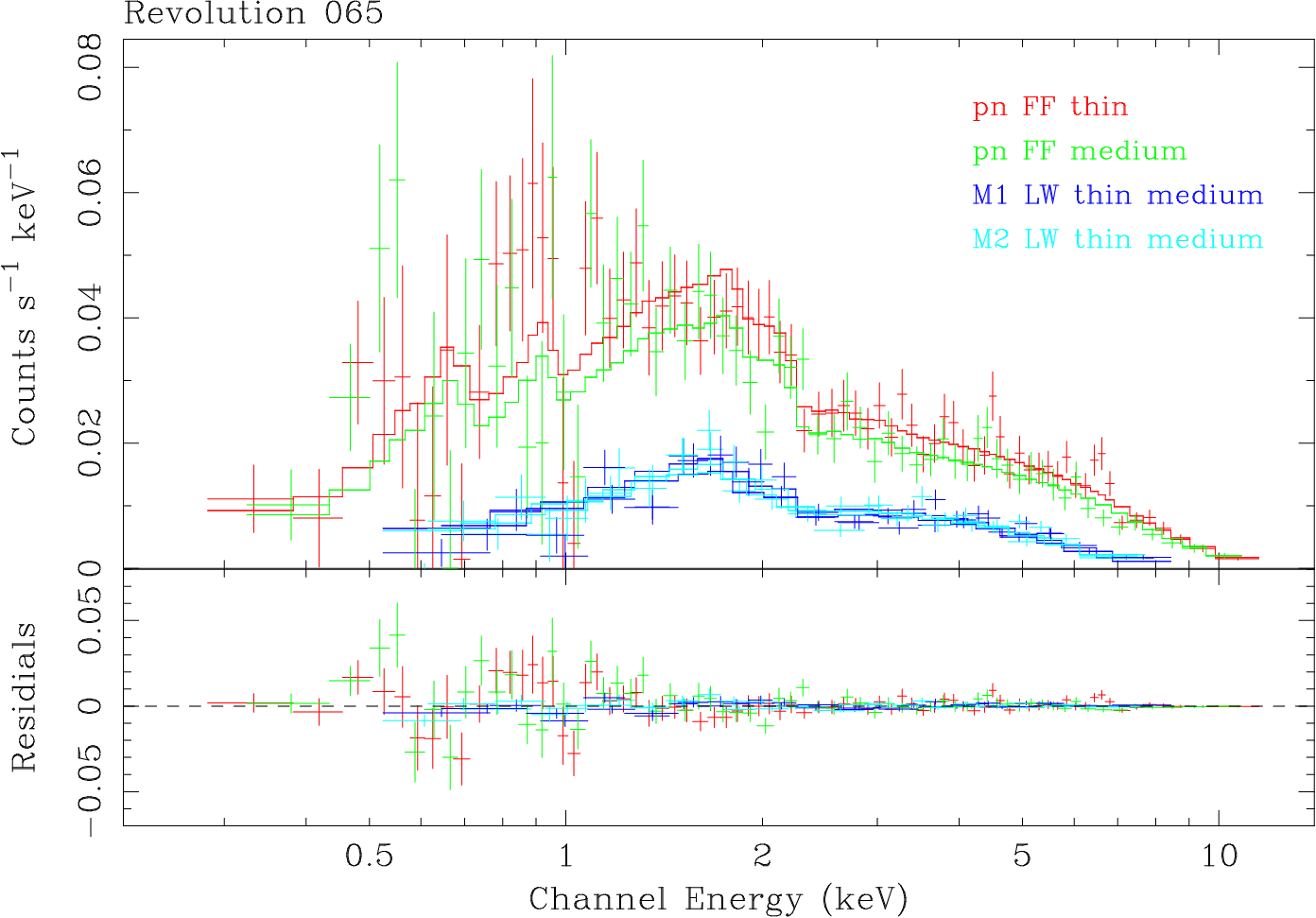}}
\hspace{3mm}
\resizebox{7.8cm}{!}{\includegraphics[angle=-90,clip=]{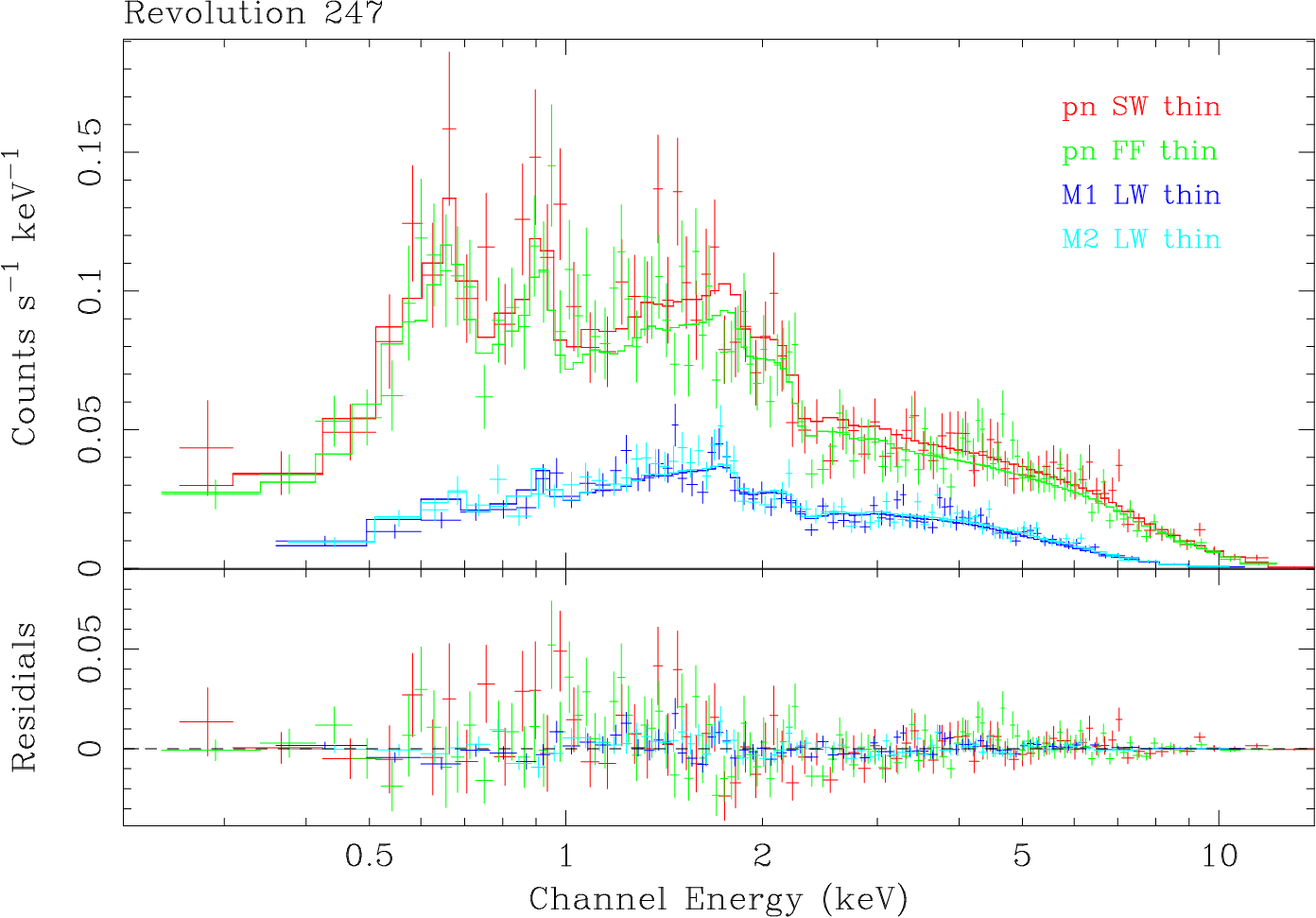}}}
\hbox{\resizebox{7.8cm}{!}{\includegraphics[angle=-90,clip=]{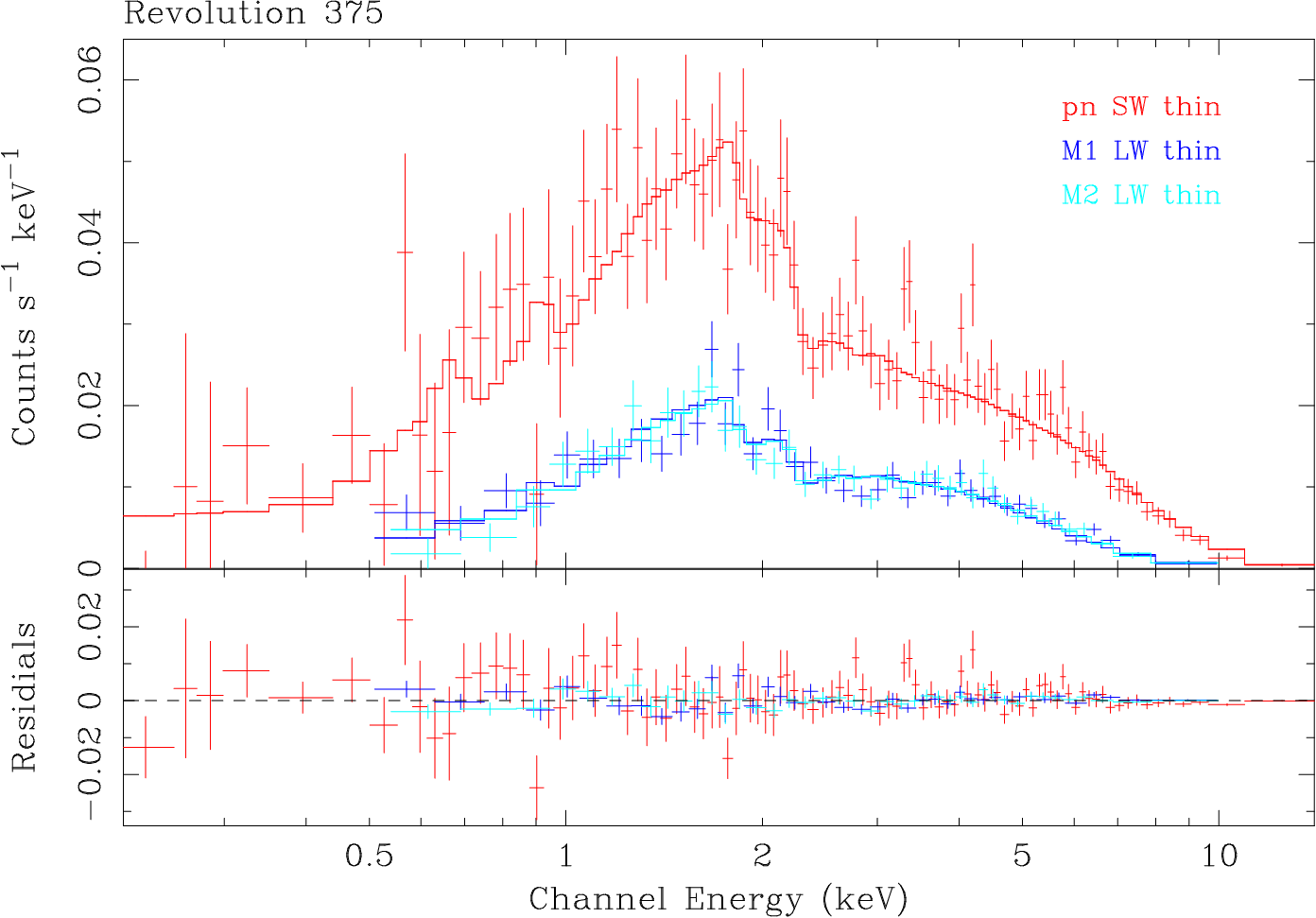}}
\hspace{3mm}
\resizebox{7.8cm}{!}{\includegraphics[angle=-90,clip=]{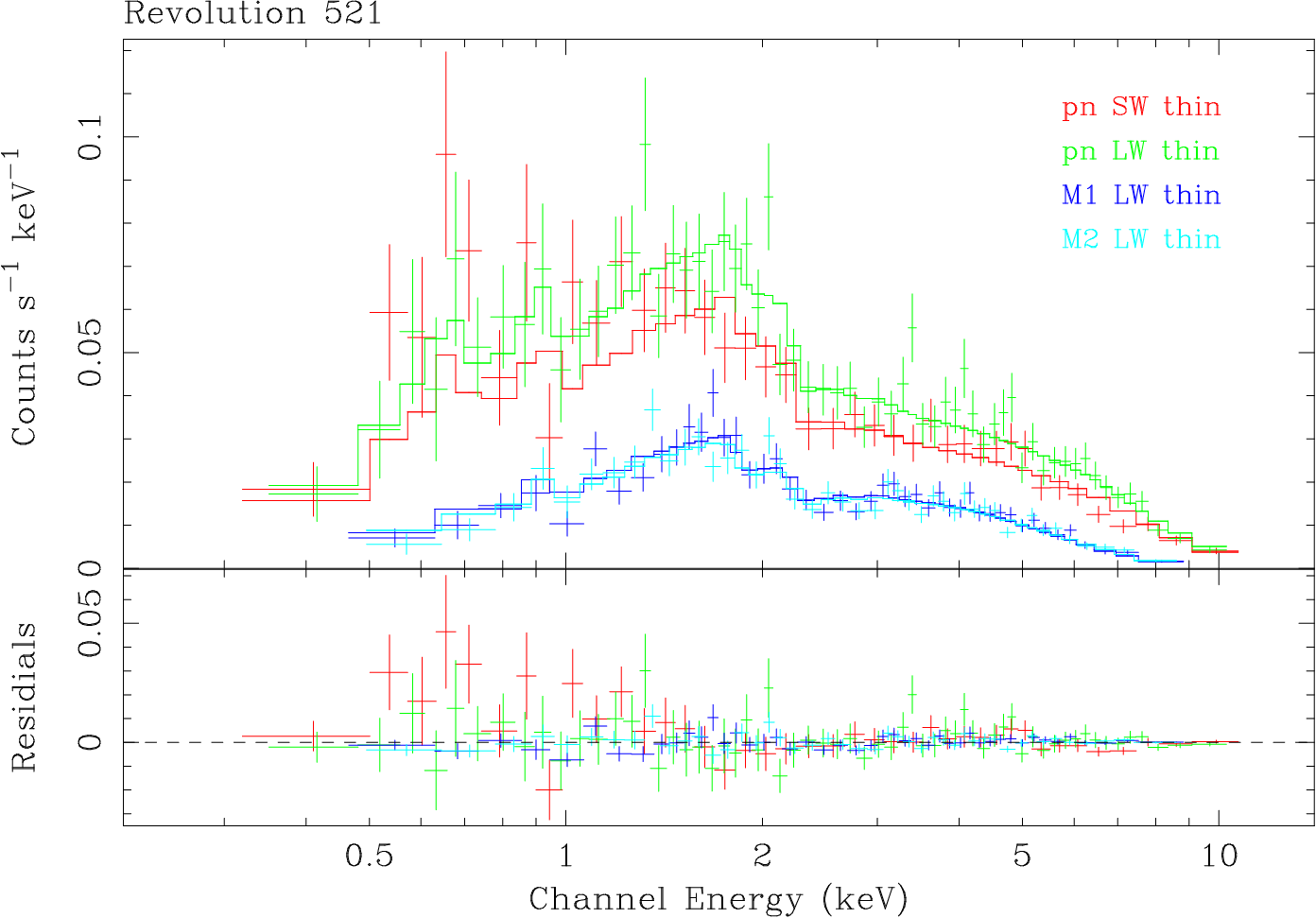}}}
\hbox{\resizebox{7.8cm}{!}{\includegraphics[angle=-90,clip=]{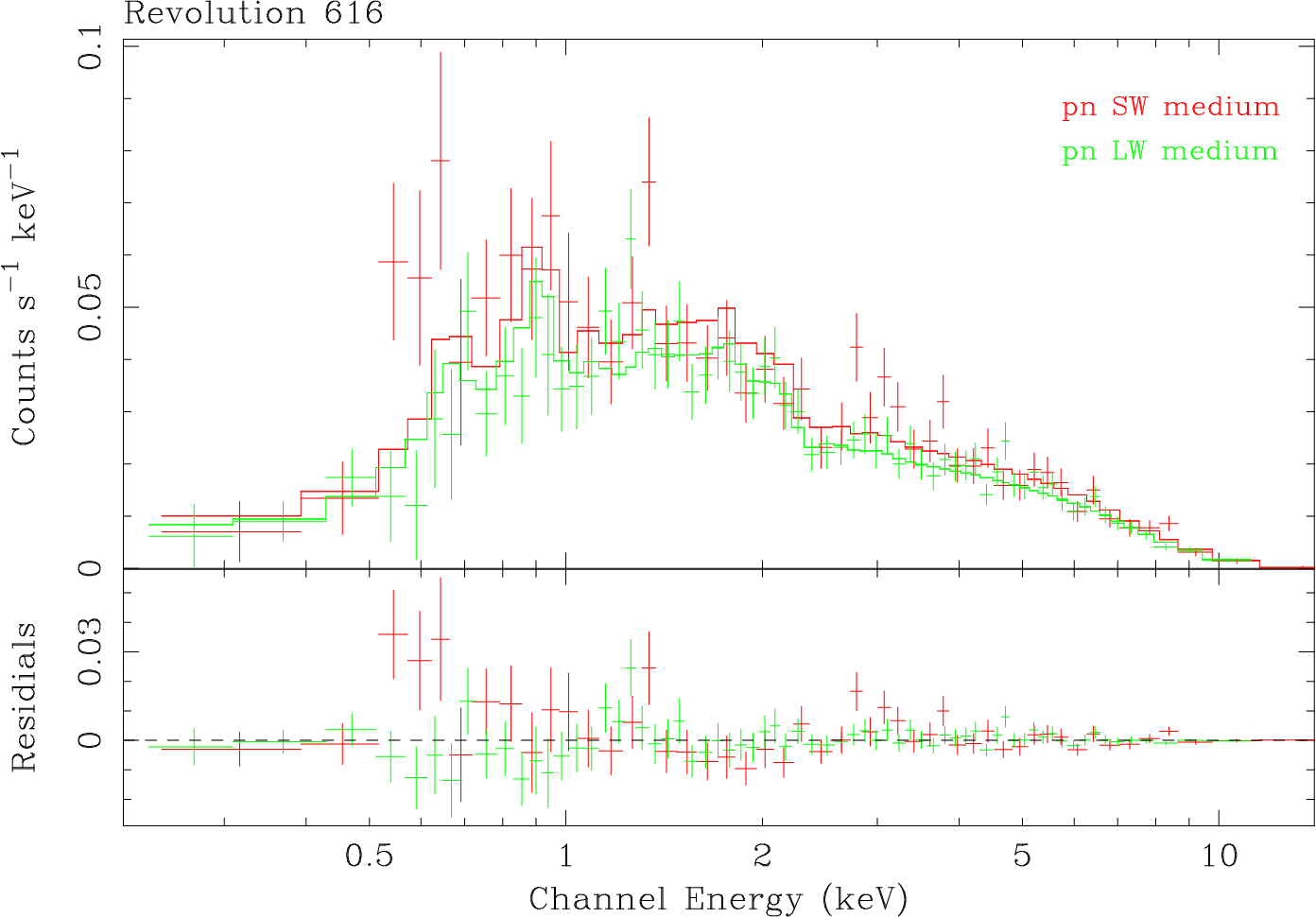}}
\hspace{3mm}
\resizebox{7.8cm}{!}{\includegraphics[angle=-90,clip=]{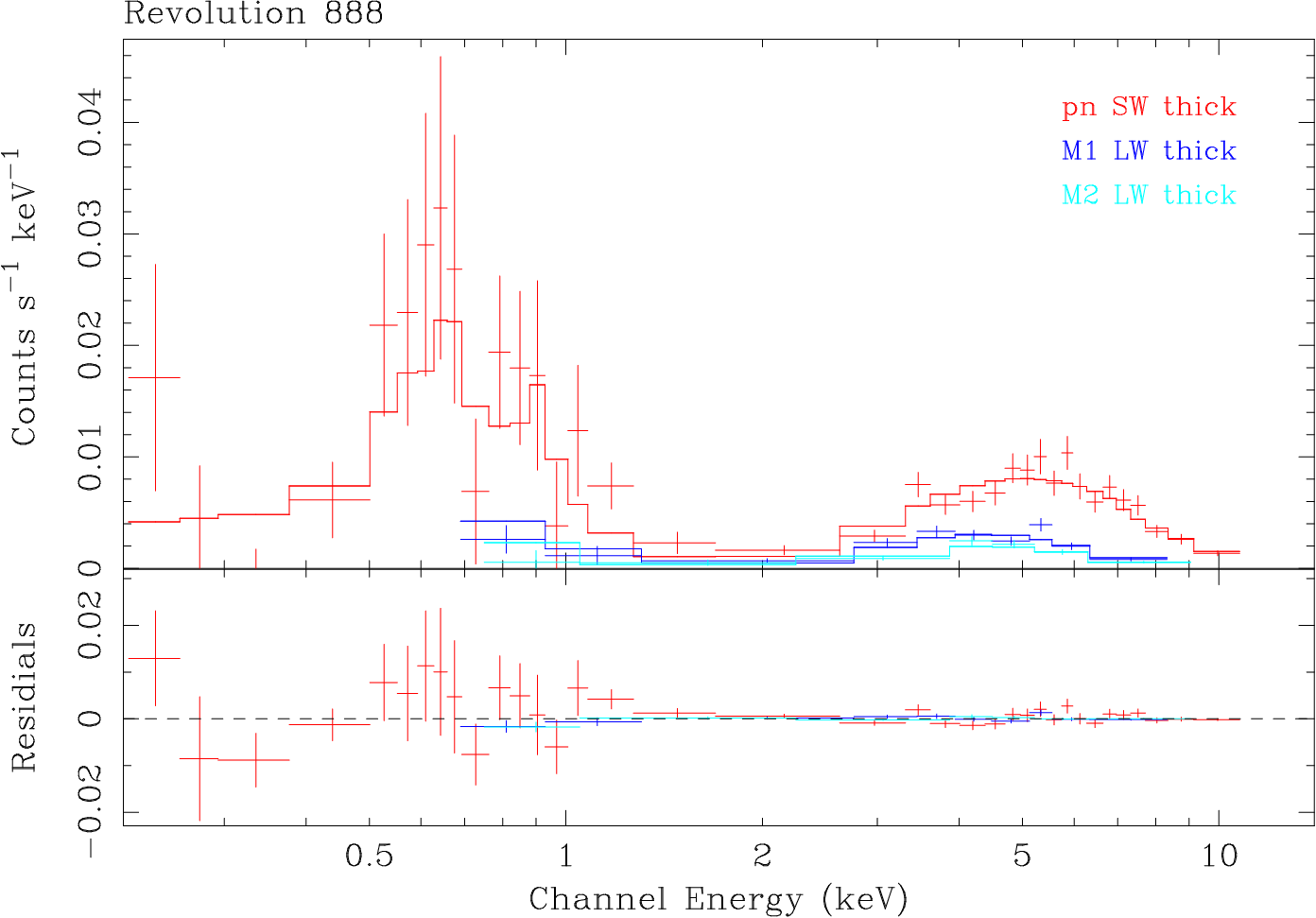}}}
\caption{EPIC spectra of \rxs\ from six observations.}
\label{epic-spectra}
\end{figure*}

\subsection{X-ray spectra}

Pulse phase averaged EPIC-pn and -MOS spectra were extracted using pattern 0$-$4 (single+double pixel) 
and pattern 0$-$12 (single $-$ quadruple) events, respectively, and binned to obtain at least 50 counts per bin. For observations split into 
parts with different readout modes (with different spectral response) separate spectra were obtained. 
Nine EPIC observations yielded a sufficiently high number of counts for a spectral analysis.
All spectra for each observation were fit simultaneously (using XSPEC v11.3) with the same model only 
allowing a free normalization factor between the different instruments and readout modes 
(which cover different time intervals with varying source intensity). 
Errors were determined for 90\% confidence levels.
A simple power-law model including photo-electric absorption by matter with solar abundance  
reproduces the data well above 1~keV while at lower energies a soft excess is apparent, 
in particular in the EPIC-pn spectra.
This soft component is clearly visible during the XMM-Newton observation of revolution 888 when 
the power-law component was highly absorbed and did not contribute below 3 keV.
The EPIC spectra of this observation were used to investigate the nature of the soft component.

To fit the soft component a black-body model and a model for a hot plasma in collisional 
equilibrium \citep[][MEKAL model in XSPEC]{1985A&AS...62..197M} 
were compared. These were used in the past to reproduce low energy excesses seen in spectra of 
SMC Be/X-ray binaries \citep{2003A&A...403..901S,2003A&A...406..471H,2004ApJ...609..133M}. 
The EPIC spectra of \rxs\
from revolution 888 were fit with these models between 0.2 and 2.0 keV where the power-law contribution
can be neglected. This results in $\chi^2$/dof values of 19.04/15 and 14.50/12 for the black-body 
and the thermal model, respectively. The thermal plasma model better reproduces the structures in the spectrum
which are probably caused by emission lines around 650 eV and 850 eV. 
Therefore, the thermal model was 
used as additional component to the power-law to fit the full EPIC spectra from all observations.

For the revolution 888 spectra two individual column density values were used for the two spectral components 
resulting in very different best fit values for \nh\ of (5.7$^{+1.9}_{-2.3}$)\hcm{21} and 107$\pm$28\hcm{21} for the 
thermal and the power-law component, respectively. For all other observations the \nh\ values were not
significantly different and therefore the same (but free to vary in the fit) value was used for both components.
Also the temperature was fixed at the value derived from the revolution 888 observation (kT = 0.15$\pm$0.04 keV). 
In Table~\ref{fit-spectra} the results for the spectral fits are summarized: 
photon index $\gamma$, absorption column density \nh, emission measure EM = $\int$ n$_{\rm H}$n$_{\rm e}$ dV,
together with fluxes and source luminosities and reduced $\chi^2$ values per degree of freedom (dof). 
In Fig.~\ref{epic-spectra} the EPIC spectra together with the best fit models are plotted.

\begin{table*}
\caption[]{XMM-Newton EPIC spectral fit results.}
\begin{center}
\begin{tabular}{lcccccccc}
\hline\hline\noalign{\smallskip}
\multicolumn{1}{l}{Instr. mode} &
\multicolumn{1}{c}{Rate$^{(1)}$} &
\multicolumn{1}{c}{\nh} &
\multicolumn{1}{c}{$\gamma$} &
\multicolumn{1}{c}{EM} &
\multicolumn{1}{c}{Flux$^{(1)}$} &
\multicolumn{1}{c}{L$_{\rm x}^{(2)}$} &
\multicolumn{1}{c}{L$_{\rm x}^{\rm i~(3)}$} &
\multicolumn{1}{c}{$\chi^2_{\rm r}$/dof} \\

\multicolumn{1}{l}{filter} &
\multicolumn{1}{c}{cts s$^{-1}$} &
\multicolumn{1}{c}{\oexpo{21}cm$^{-2}$} &
\multicolumn{1}{c}{} &
\multicolumn{1}{c}{10$^{58}$ cm$^{-3}$} &
\multicolumn{1}{c}{erg cm$^{-2}$ s$^{-1}$} &
\multicolumn{1}{c}{erg s$^{-1}$} &
\multicolumn{1}{c}{erg s$^{-1}$} &
\multicolumn{1}{c}{} \\

\noalign{\smallskip}\hline\noalign{\smallskip}
\multicolumn{9}{l}{2000 Apr. 16/17 (Revolution 065)} \\
 PN FF thin   &  0.177 & 6.5$\pm$1.5        & 0.88$\pm$0.07 & 6.93 & 1.61\expo{-12} & 6.87\expo{35} & 2.00\expo{36} & \\
 M1 LW thin   &  0.052 &                    &		    & 7.18 & 1.66\expo{-12} & 7.13\expo{35} & 2.08\expo{36} & \\
 M2 LW thin   &  0.051 &                    &		    & 6.90 & 1.60\expo{-12} & 6.84\expo{35} & 2.00\expo{36} & \\
 PN FF medium &  0.156 &                    &		    & 6.10 & 1.41\expo{-12} & 6.04\expo{35} & 1.76\expo{36} & \\
 M1 LW medium &  0.048 &                    &		    & 6.56 & 1.52\expo{-12} & 6.50\expo{35} & 1.90\expo{36} & \\
 M2 LW medium &  0.049 &                    &		    & 6.63 & 1.54\expo{-12} & 6.57\expo{35} & 1.92\expo{36} & 1.29/209 \\
\noalign{\smallskip}							    	   	 	     
\multicolumn{9}{l}{2001 Apr. 14/15 (247)} \\
 PN SW thin   &  0.413 & 6.2$\pm$0.4	     & 0.84$\pm$0.03 & 22.1 & 3.59\expo{-12} & 1.54\expo{36} & 5.61\expo{36} & \\
 PN FF thin   &  0.376 &		     &  	     & 20.0 & 3.25\expo{-12} & 1.39\expo{36} & 5.08\expo{36} & \\
 M1 LW thin   &  0.121 &		     &  	     & 21.9 & 3.55\expo{-12} & 1.52\expo{36} & 5.55\expo{36} & \\
 M2 LW thin   &  0.126 &		     &  	     & 21.9 & 3.57\expo{-12} & 1.53\expo{36} & 5.58\expo{36} & 1.24/347 \\
\noalign{\smallskip}							    	   	 	     
\multicolumn{9}{l}{2001 Dec. 25/26 (375)} \\
 PN SW thin   &  0.189 & 7.0$\pm$1.6        & 0.90$\pm$0.08 & 6.03 & 1.77\expo{-12} & 7.56\expo{35} & 1.94\expo{36} & \\
 M1 LW thin   &  0.064 &                    &	            & 6.86 & 2.01\expo{-12} & 8.61\expo{35} & 2.20\expo{36} & \\
 M2 LW thin   &  0.063 &                    &	            & 6.73 & 1.97\expo{-12} & 8.44\expo{35} & 2.16\expo{36} & 1.19/177 \\
\noalign{\smallskip}							    	   	 	     
\multicolumn{9}{l}{2002 Oct. 13    (521)} \\
 PN SW thin   &  0.244 & 6.8$\pm$1.0        & 0.85$\pm$0.05 & 12.1 & 2.35\expo{-12} & 1.01\expo{36} & 3.27\expo{36} & \\
 PN LW thin   &  0.289 &                    &		    & 14.7 & 2.88\expo{-12} & 1.23\expo{36} & 4.00\expo{36} & \\
 M1 LW thin   &  0.096 &                    &		    & 15.4 & 3.00\expo{-12} & 1.28\expo{36} & 4.16\expo{36} & \\
 M2 LW thin   &  0.094 &                    &               & 15.0 & 2.91\expo{-12} & 1.24\expo{36} & 4.04\expo{36} & 1.08/185 \\
\noalign{\smallskip}							    	   	 	     
\multicolumn{9}{l}{2002 Dec. 14    (552)} \\
 PN SW thin   &  0.244 & 6.5$\pm$1.3        & 0.91$\pm$0.09 & 4.83 & 5.50\expo{-13} & 2.36\expo{35} & 1.12\expo{36} & \\
 PN LW thin   &  0.289 &                    &		    & 5.74 & 6.56\expo{-13} & 2.81\expo{35} & 1.33\expo{36} & \\
 M1 LW thin   &  0.096 &                    &		    & 5.63 & 6.41\expo{-13} & 2.75\expo{35} & 1.30\expo{36} & \\
 M2 LW thin   &  0.094 &                    &		    & 5.83 & 6.64\expo{-13} & 2.84\expo{35} & 1.35\expo{36} & 0.82/63 \\
\noalign{\smallskip}							    	   	 	     
\multicolumn{9}{l}{2003 Apr. 20/21 (616)} \\
 PN SW medium &  0.200 & 8.4$\pm$1.2        & 0.96$\pm$0.06 & 30.9 & 1.88\expo{-12} & 8.04\expo{35} & 6.40\expo{36} & \\
 PN LW medium &  0.166 &                    &	            & 26.9 & 1.63\expo{-12} & 6.98\expo{35} & 5.55\expo{36} & 0.99/111 \\
\noalign{\smallskip}							    	   	 	     
\multicolumn{9}{l}{2003 Nov. 16    (721)} \\
 PN FF thick  &  0.143 & 6.2$\pm$1.1        & 0.91$\pm$0.05 & 8.53 & 1.32\expo{-12} & 5.65\expo{35} & 2.14\expo{36} & \\
 M1 LW thin   &  0.050 &                    &		    & 9.55 & 1.48\expo{-12} & 6.32\expo{35} & 2.40\expo{36} & \\
 M2 LW thin$^{(4)}$ & 0.036 &               &		    & 7.66 & 1.18\expo{-12} & 5.07\expo{35} & 1.92\expo{36} & 1.48/125 \\
\noalign{\smallskip}							    	   	 	     
\multicolumn{9}{l}{2004 Apr. 28    (803)} \\
 PN LW thick  &  0.030 & 6.6$\pm$1.7        & 1.12$\pm$0.15 & 3.28 & 2.82\expo{-13} & 1.21\expo{35} & 7.21\expo{35} & 0.72/25  \\
\noalign{\smallskip}							    	   	 	     
\multicolumn{9}{l}{2004 Oct. 14    (888)$^{(5)}$} \\
 PN LW thick  &  0.055 & 107$\pm$28 & 1.03$\pm$0.34 & 4.19 & 7.31\expo{-13} & 3.13\expo{35} & 1.34\expo{36} & \\
 M1 LW thick  &  0.013 &            &               & 4.40 & 7.67\expo{-13} & 3.28\expo{35} & 1.40\expo{36} & \\
 M2 LW thick$^{(4)}$ & 0.008 &      &               & 3.18 & 5.56\expo{-13} & 2.38\expo{35} & 1.02\expo{36} & 1.48/45  \\
\noalign{\smallskip}\hline
\end{tabular}
\end{center}
$^{(1)}$ Net count rates and observed total fluxes in the 0.2$-$10.0 keV band.\\
$^{(2)}$ X-ray 0.2$-$10.0 keV luminosity (including absorption) for a distance of 60 kpc. \\ 
$^{(3)}$ Source intrinsic X-ray luminosity in the 0.2$-$10.0 keV band (corrected for absorption).\\
$^{(4)}$ Source on bad CCD column.\\
$^{(5)}$ Individual column density for the thermal plasma component \nh\ = (5.7$^{+1.9}_{-2.3}$)\hcm{21}. The temperature of the thermal plasma
model component was determined from this observation to kT = 0.15$\pm$0.04 keV and fixed in the fits to the other observations.

\label{fit-spectra}
\end{table*}

\subsection{Energy dependence of pulsations}

\begin{figure}
\resizebox{8.5cm}{!}{\includegraphics[angle=-90,clip=]{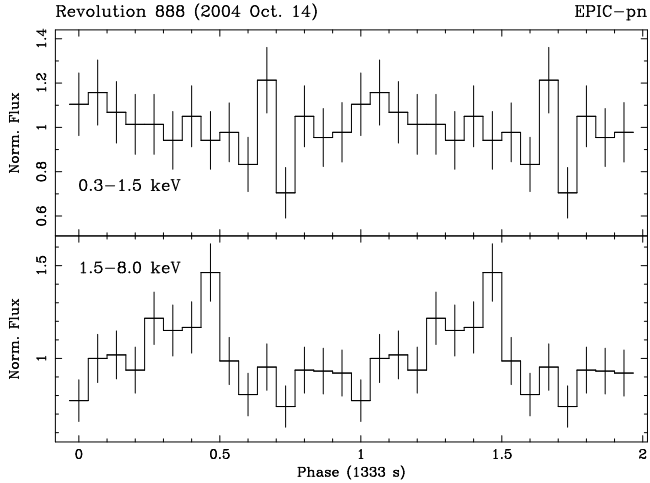}}
\caption{EPIC-pn light-curve from revolution 888 in soft and hard energy bands folded with the pulse period.}
\label{pulse888}
\end{figure}

The pulsed fraction observed from \rxs\ during revolution 616 is largely reduced (if present at all) 
in the 0.3$-$1.5 keV band compared to the 1.5$-$8.0 keV band. However, the observation from 
revolution 521 shows that there can be still a large contribution
from the power-law component in the soft band. The observation from revolution 888 where the soft 
component was completely isolated is ideally suited to investigate the pulse profiles of both spectral 
components separately. This is shown in Fig.~\ref{pulse888} and although the statistics is relatively poor,
pulsations can be detected above 1.5 keV while below they are strongly reduced or may not be present at all.

\subsection{Long-term intensity variations}

\begin{figure}
\resizebox{8.0cm}{!}{\includegraphics[angle=-90,clip=]{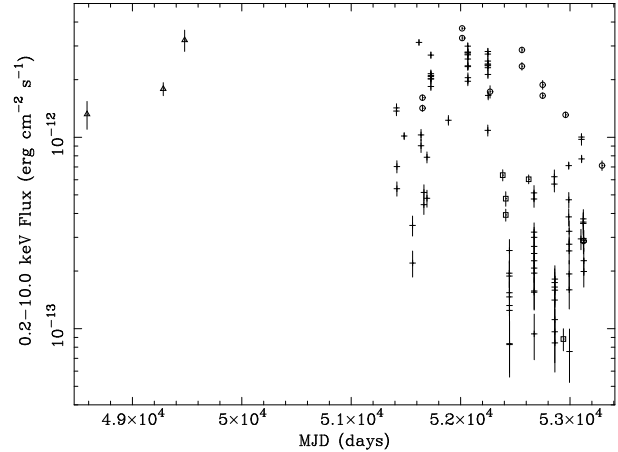}}
\caption{X-ray flux variations from \rxs\ since the first ROSAT (triangles) observation in
Nov./Dec. 1991. Circles mark XMM-Newton observations with fluxes directly derived from
spectral fits to the EPIC-pn spectra (which yield two data points when two read-out modes 
were used. All other fluxes were converted from count rates assuming a power-law + MEKAL
model with typical parameters as derived from the EPIC spectra (see text). Crosses indicate 
Chandra observations and squares XMM-Newton observation with the source at low
intensity.}
\label{flux-hist}
\end{figure}

To look at intensity variations of \rxs\ on time scales of days to years we converted
ROSAT and Chandra count rates obtained from source detection analyses into 0.2$-$10.0 keV fluxes.
We assumed a power-law plus MEKAL spectrum with photon index 0.9, column density of 6.5\hcm{21}
and temperature 0.15 keV, typical values as derived from the EPIC spectra. 
Similarly, for EPIC observations with insufficient statistical
quality (short exposure and/or low source flux) count-rates were used. For the estimates of ROSAT 
fluxes relatively large systematic errors may be introduced due to the strong influence of the assumed 
column density on the soft 0.1$-$2.4 keV band count rate to luminosity conversion (an additional \nh\ of 1.0\hcm{21}
changes the conversion factor by 12\%). The measured fluxes as 
function of time over nearly 13 years are plotted in Fig.~\ref{flux-hist}. There seems to be a base 
flux level of $\sim$7.6\ergcm{-13} below which the source was never seen. The maximum flux of 
3.7\ergcm{-12} is a factor of 50 higher and was reached during the XMM-Newton observation of 
revolution 247 (second part with EPIC-pn in FF mode). A similar outburst level was seen also 
during the third ROSAT observation, four Chandra and one other XMM-Newton observation. Chandra 
observations were often performed in blocks spread over $\sim$0.7 days during which the source 
stayed at a similar intensity level, indicating that high and low states last longer than that. 
The source was generally brighter (both, low and high intensity levels decreased) 
during the first half as compared to the second half of the last 
five years (Fig.~\ref{flux-hist}) which may be related 
to long-term changes in the properties of the circum-stellar matter around the mass donor star.

\begin{table*}
\caption[]{Optical identifications.}
\begin{center}
\begin{tabular}{lcccccc}
\hline\hline\noalign{\smallskip}
\multicolumn{1}{c}{Colours (catalogue)} &
\multicolumn{1}{c}{R.A. and Dec. (2000.0)} &
\multicolumn{1}{c}{Vmag} &
\multicolumn{1}{c}{B$-$V} &
\multicolumn{1}{c}{U$-$B} &
\multicolumn{1}{c}{V$-$R} &
\multicolumn{1}{c}{V$-$I} \\

\noalign{\smallskip}\hline\noalign{\smallskip}
UBVR \citep{2002ApJS..141...81M} & 01$^{\rm h}$03$^{\rm m}$37\fs46 \hspace{1mm}$-$72\degr01\arcmin33\farcs2 & 14.54 & $-$0.06 & $-$0.99 & $-$0.08 & --      \\
UBVI \citep{2002AJ....123..855Z} & 01$^{\rm h}$03$^{\rm m}$37\fs52 \hspace{1mm}$-$72\degr01\arcmin33\farcs0 & 14.65 & $-$0.11 & $-$1.08	& --	  & $-$0.04 \\
\noalign{\smallskip}\hline
\end{tabular}
\end{center}
\label{tab-ids}
\end{table*}

\subsection{Source identification}

X-ray source positions were obtained from the analysis of all observations
where the source was detected in the ACIS-I CCDs.
The average position from 71 observations with a standard deviation of 1\farcs1
is only 0\farcs07 away from the 2MASS object 
01033752$-$7201330 at RA = 01$^{\rm h}$03$^{\rm m}$37\fs524 and Dec = --72\degr01\arcmin33\farcs03
(J2000.0) for which an error of 0\farcs11 is given \citep{2003yCat.2246....0C}. For comparison the
average EPIC position is within 0\farcs8 of the 2MASS position.
This star is covered by the SMC UBVR CCD survey of \citet{2002ApJS..141...81M} and also listed in 
the UBVI photometry catalogue of \citet{2002AJ....123..855Z}. Optical positions and 
available magnitudes are summarized in Table~\ref{tab-ids}. 
The star can be identified with the emission line object
\citep[\ma, ][]{1993A&AS..102..451M}.
The presence of a $\sim$14.6 Vmag star inside the error circle of the Chandra position 
clearly establishes it as the optical counterpart of the X-ray source.

The inferred effective temperature and bolometric luminosity of the optical counterpart given by
\citet{2002ApJS..141...81M} suggest a spectral type of O5 V. Assuming this spectral type 
the measured B$-$V index imply E(B$-$V) of 0.32 and \nh\ = 1.9\hcm{21}.

\section{Discussion}

\subsection{\rxs, a Be/X-ray binary pulsar}

The discovery of a 1323 s modulation in the X-ray flux from \rxs\ increases the number of pulsars in 
the SMC to 49 from which up to 46 are most likely neutron stars in HMXB systems. The pulse period of
1323 s is the longest known from SMC pulsars. It considerably exceeds that of AX\,J0049.4$-$7323 (= RX\,J0049.7$-$7323)
\citep[756 s; ][]{2003PASJ...55..161Y} and is similar to that of \rxl\ \citep{1998A&A...335..587H} in the Milky Way.
Optical spectra of 33 counterparts from HMXBs in the SMC identify the majority as Be-star binaries 
\citep{2005MNRAS.356..502C}, while only one (SMC\,X-1) is a supergiant
system. Brightness and colours of the optical counterpart and the presence of \Halp\
emission (MA93) strongly suggest that also \rxs\ is a Be/X-ray binary system.

The X-ray spectrum of \rxs\ is typical for Be/X-ray binaries in the SMC. The power-law photon index of 
$\sim$0.9 is well within the relatively narrow range observed from these systems \citep{2004A&A...414..667H}. 
The absorbing column density was found to be similar during most observations yielding (6$-$9)\hcm{21},
except during the XMM-Newton observation in revolution 888 when \nh\ exceeded \ohcm{23}. Such high values 
are usually seen only from supergiant HMXBs when the X-ray source is viewed through the dense innermost parts 
of the stellar wind of the supergiant before or after eclipse 
\citep[e.g. 4U1700$-$37 and Vela\,X-1;][]{1989ApJ...343..409H,1999ApJ...525..921S}. This suggests that we may view
the system of \rxs\ at high inclination and during 2004, Oct. 14 (revolution 888) 
the line of sight to the X-ray source crossed the wind region very close to the Be star. 
Since Be/X-ray binaries are much wider systems in comparison to supergiant systems such cases should be very rare.
Indeed, \rxs\ is to our knowledge the only Be/X-ray binary which has shown a spectrum with such extreme absorption.
Alternatively absorption by dense material close to the neutron star cannot be ruled out, but may be expected to occur
more often.

\subsection{Soft emission in the X-ray spectra of Be/X-ray binaries}

\begin{figure}
\resizebox{8.5cm}{!}{\includegraphics[angle=-90,clip=]{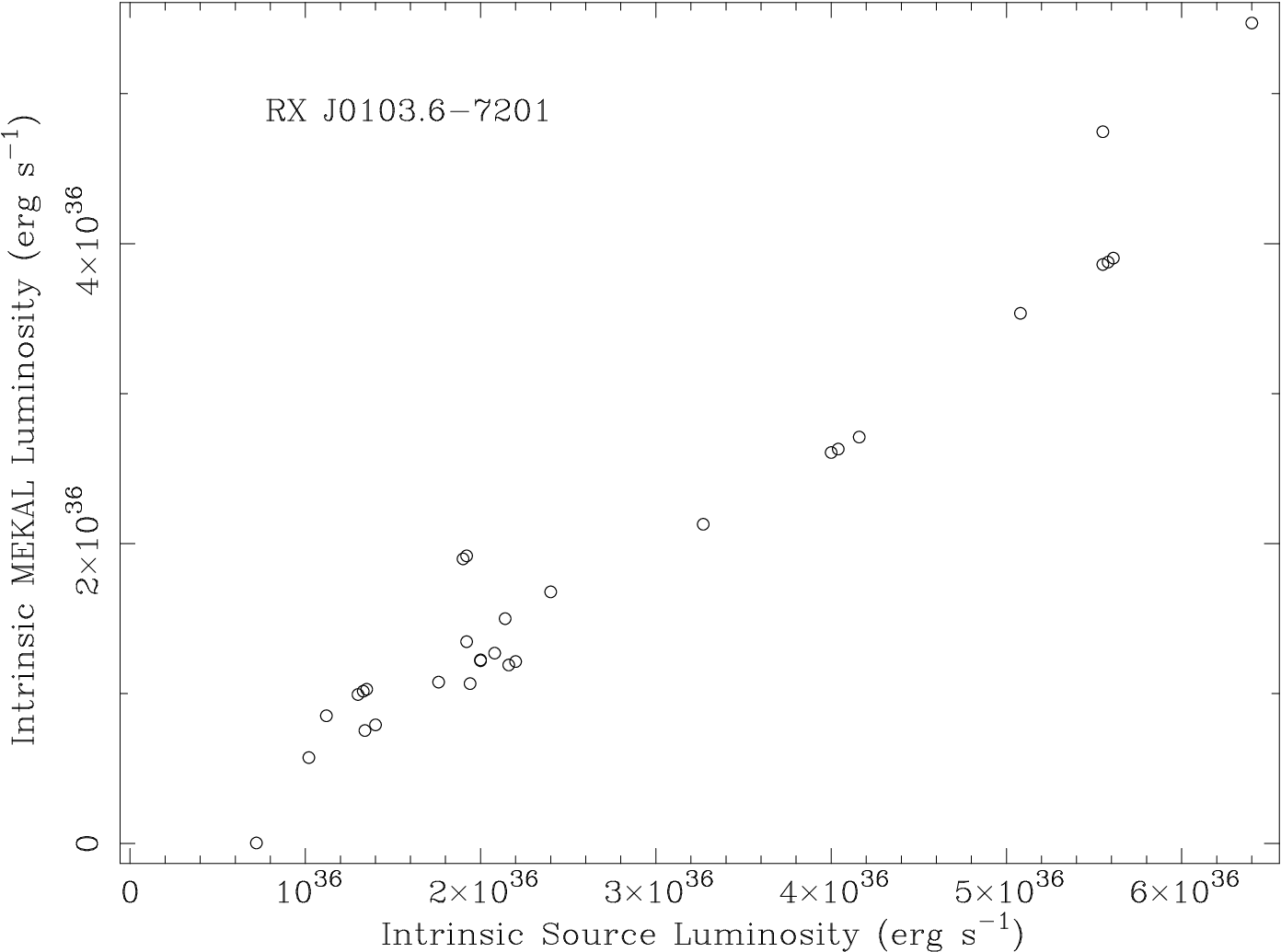}}
\vspace{5mm}
\resizebox{8.5cm}{!}{\includegraphics[angle=-90,clip=]{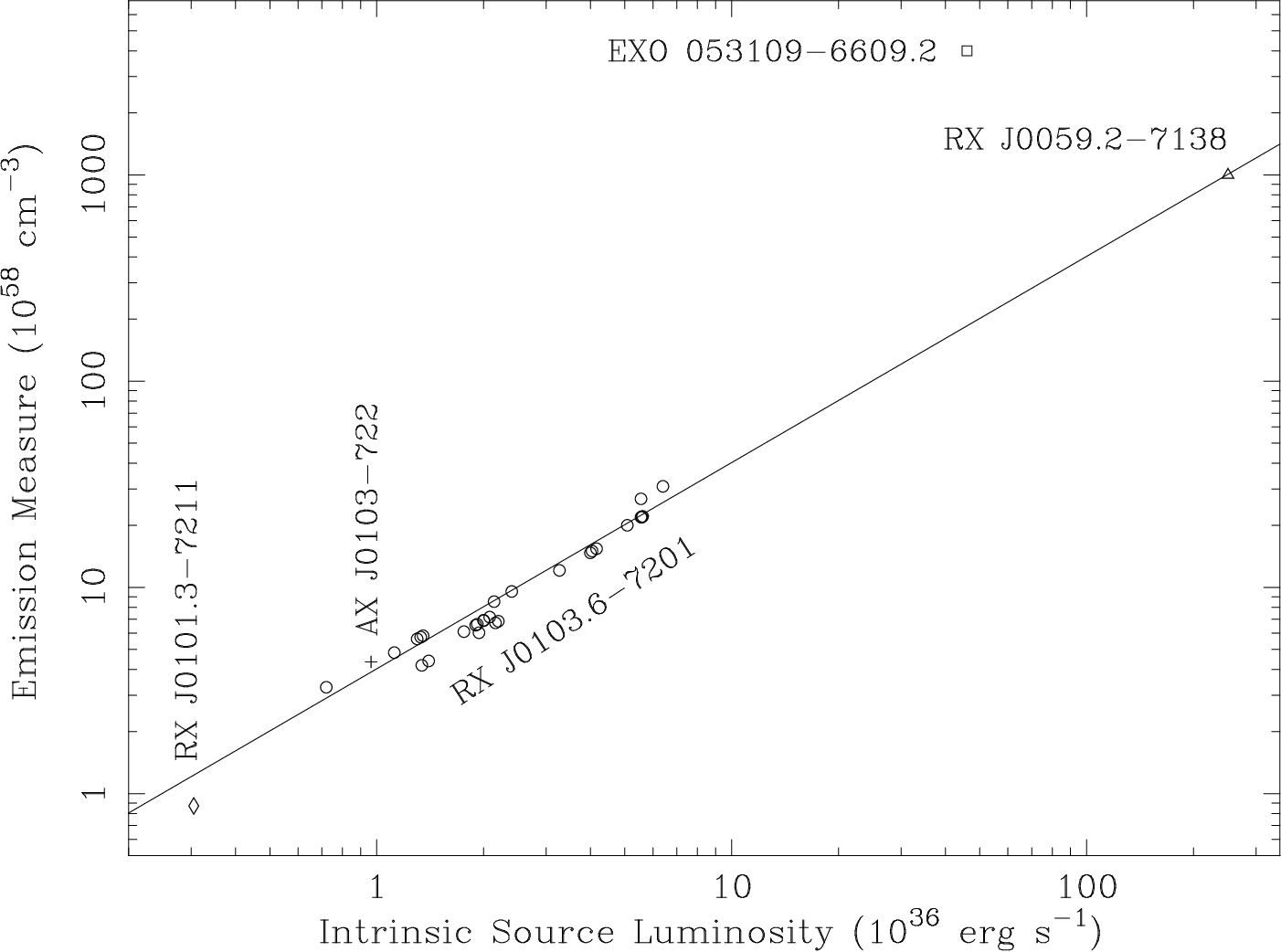}}
\caption{Top: Luminosity of the soft spectral component (modeled as MEKAL component) 
in the EPIC spectra of \rxs\ vs. total source luminosity. Both luminosities are source 
intrinsic, i.e. corrected for photo-electric absorption.
Bottom: Emission measure of the thermal plasma component as function of the total source intrinsic 
luminosity including results from other Be/X-ray binaries.}
\label{em-measure}
\end{figure}

A soft spectral component in the X-ray spectrum of \rxs\ was first reported by \citet{2003A&A...403..901S} who 
analyzed the XMM-Newton data from revolution 247 and fit the EPIC-pn spectrum with thermal plasma and power-law 
components. This soft emission became clearly visible in the revolution 888 observation when 
both spectral components were separated into different energy bands.
From Table~\ref{fit-spectra} it is evident that the intensity of the soft component, given as emission measure 
(which directly reflects the intrinsic luminosity of the soft component)
in the range of (3$-$30)\expo{58} cm$^{-3}$ strongly correlates with the intrinsic source luminosity.
This is demonstrated in the top panel of Fig.~\ref{em-measure} where the luminosity of the soft component is
plotted versus the total source luminosity.

Similar soft emission was also observed from the pulsars \exo\ with XMM-Newton \citep{2003A&A...406..471H} and 
RX\,J0059.2$-$7138 in a simultaneous ROSAT/ASCA observation \citep{2000PASJ...52..299K} with emission measures 
of $\sim$4\expo{61} cm$^{-3}$ and $\sim$\oexpo{61} cm$^{-3}$, respectively. 
RX\,J0059.2$-$7138 was observed at a luminosity of 2.5\ergs{38}, i.e. both luminosity and emission measure
were a factor of 40 higher compared to the maximum values observed from \rxs.
On the other hand, for \exo\ an intrinsic source luminosity of 4.6\ergs{37} was reported which yields a ratio 
EM[cm$^{-3}$]/L$_{\rm x}$[erg s$^{-1}$] of 8.7\expo{23} a factor of 20 higher than for \rxs. 
\citet{2003A&A...403..901S} report two more SMC pulsars which exhibit a soft emission component. We re-analyzed
the EPIC-pn spectra of RX\,J0101.3$-$7211 and AX\,J0103$-$722 in a consistent way as was done for \rxs. Keeping the
temperature of the thermal plasma component fixed at 0.15 keV and using a common \nh\ for soft and hard (power-law)
component we derive EM and intrinsic source luminosity (0.2$-$10.0 keV). The results, together with the values 
found in the literature, are included in the bottom panel of Fig.~\ref{em-measure}.
They are consistent with the linear relation between EM (or luminosity of the soft component) and total source 
luminosity derived for \rxs\ with EM [cm$^{-3}$] = 4\expo{22}$\times$L$_{\rm x}$ [erg s$^{-1}$] 
(plotted as straight line). Here we use EM instead of luminosity of the soft component to be able to compare 
with published data.

\citet{2004ApJ...614..881H} explored four possible mechanisms for the origin of the soft excess in X-ray pulsars:
1) re-processing of hard X-rays by optically thick, dense material near the neutron star; 2) re-processing by 
optically thin material; 3) thermal emission from collisionally ionized plasma and 4) soft emission from the 
accretion column and the neutron star surface. They conclude that for luminous sources 
(L$_{\rm x}>10^{38}$ erg s$^{-1}$) only case 1) can explain the strong soft component while for luminosities 
below \oergs{36} all three other cases can contribute.

The tight correlation between the luminosity of the soft spectral component and the total source luminosity
we found for \rxs\ over a factor of 10 in luminosity and 
moreover extending to three orders of magnitude when including results from three
other SMC Be/X-ray binary pulsars strongly suggests that there is a single mechanism at work over a wide range
of source intensity. The correlation as such makes emission from a collisionally ionized plasma unlikely
and supports the models involving re-processing of the X-ray emission from the neutron star. Therefore, the MEKAL
model is actually not the proper model to derive parameters of the emitting plasma like temperatures 
or elemental abundances. However, it best fits the soft part of the EPIC spectra among all available simple 
XSPEC models and allows to accurately determine fluxes.

The pulsed fraction observed from \rxs\ is largely reduced (if present at all) at energies below 1.5 keV. 
Similarly, no pulsations at energies below 0.4 keV were seen from the 2.76 s pulsar RX\,J0059.2$-$7138 
\citep{2000PASJ...52..299K} which indicates a size for a re-processing region with
r $>$ c$\times$P$_{\rm spin}\sim 10^{11}$ cm. To smear out longer pulse periods the size must be correspondingly
larger and for \rxs\ would require a size $>$ 4\expo{13} cm. 
An independent very rough estimate for the size of the re-processing region can be derived from the
emission measure $\int$~n$_{\rm H}$n$_{\rm e}$ dV $\sim$ 4$\pi$/3 r$^3$$\times$n$_{\rm H}^2$ and the absorbing 
column density $\int$~n$_{\rm H}$ dr $\sim$ r$\times$\nh. For EM in the range of (3$-$30)\expo{58} cm$^{-3}$
and a typical local \nh\ of 6\hcm{21}, r is in the range 2\expo{14} cm to 2\expo{15} cm (increasing
with higher EM and hence source luminosity). If the large absorption
observed from revolution 888 for the power-law component is caused by dense stellar matter close to the 
Be star, than these latter size estimates must be regarded as upper limit. A lower limit of 1\expo{12} cm
can then be derived using the high \nh\ value. 
These simple estimates argue for reprocessing of the hard X-rays in a large part of the optically thin
stellar wind. The correlation of soft flux with source intensity for four Be/X-ray binaries in the SMC
suggests that this mechanism is the dominant process to create the soft spectral component at least in 
HMXBs with a Be star as mass donor.

It remains to be explained why \exo\ exhibits a much stronger soft component compared to the SMC
Be/X-ray pulsars. It is unlikely that the higher metallicity of the LMC can cause a factor of
20 increase in reprocessing efficiency. \exo\ has shown a more complex spectrum during the XMM-Newton
observation reported by \citet{2003A&A...406..471H} which may be interpreted as partial covering
of the X-ray source. Hence, a large fraction of the hard emission might be completly blocked
and the total source luminosity underestimated. Other Be/X-ray binaries in the LMC with a soft 
emission component need to be observed to verify if \exo\ is an exceptional case 
\citep[e.g. A\,0538$-$66, ][]{1993A&A...274..304M}. 

%

\begin{acknowledgements}
We are grateful to Manami Sasaki for providing the EPIC spectra of SMC pulsars with soft 
emission for our re-analysis. The XMM-Newton project is supported by the Bundesministerium 
f\"ur Bildung und For\-schung / Deutsches Zentrum f\"ur Luft- und Raumfahrt (BMBF / DLR), the
Max-Planck-Gesellschaft and the Heidenhain-Stif\-tung. 
\end{acknowledgements}

\bibliographystyle{aa}
\bibliography{mcs,hmxb,general,myrefereed,myunrefereed,mytechnical}

\end{document}